\documentclass[%
superscriptaddress,
 amsmath,amssymb,
 aps,
 prl,
 letterpaper,
]{revtex4-2}

\usepackage{tikz}
\usepackage[english]{babel}
\usetikzlibrary{trees}
\usetikzlibrary{decorations.pathmorphing}
\usetikzlibrary{decorations.markings}
\usepackage{tikz-cd}
\usetikzlibrary{arrows,arrows.meta}

\tikzset{
    photon/.style={decorate, decoration={snake,segment length=1.5mm}, draw=black},
    electron/.style={draw=black, postaction={decorate},
        decoration={markings,mark=at position .55 with {\arrow[draw=black]{>}}}},
    gluon/.style={decorate, draw=magenta,
        decoration={coil,amplitude=4pt, segment length=5pt}},
    boundelectron/.style={thick, double}
}

\usepackage{hyperref}
\hypersetup{
    colorlinks=true,
    linkcolor=black,
    filecolor=black,      
    urlcolor=blue,
    citecolor=black,
}

\usepackage{graphicx}
\usepackage{siunitx}
\usepackage{dcolumn}
\newcolumntype{.}{D{.}{.}{8}}
\usepackage{bm}
\usepackage{rotating}
\usepackage{comment}
\usepackage{empheq}

\usepackage[utf8]{inputenc}
\usepackage{physics}
\usepackage{amsmath, amssymb, amsthm}
\usepackage{tabularx,booktabs,array,dcolumn}
\usepackage{setspace}
\usepackage{vmargin}
\usepackage{xcolor}
\usepackage{graphicx,psfrag,subfigure}

\usepackage{float}
\usepackage[all]{xy}
\usepackage{verbatim}
\usepackage{dsfont}
\usepackage{empheq}

\usepackage{slashed}
\usepackage{multirow}

\newcommand{\bos}[1]{\boldsymbol{#1}}
\newcommand{\mr}[1]{\mathrm{#1}}

\def\m0{m^{(0)}}

\def\iim{\mr{i}}
\def\eem{\mr{e}}

\def\iim{\mr{i}}
\def\eem{\mr{e}}
\def\dd{\mr{d}}
\def\grad{\bos{\nabla}}
\def\nb{n_\text{b}}

\def\bk{\bos{k}}

\def\bp{\bos{p}}

\def\br{\bos{r}}

\def\bx{\bos{x}}
\def\by{\bos{y}}

\def\balpha{\boldsymbol{\alpha}}

\def\grad{\boldsymbol{\nabla}}

\def\he2p{He$_2^+$}
\def\h2p{H$_2^+$}

\newcolumntype{d}[1]{D{.}{.}{#1}}

\begin{document}

\title{Gaussian basis set approach to one-loop self-energy}
\author{D\'avid Ferenc} 
\email{david.ferenc@irsamc.ups-tlse.fr}
\affiliation{Laboratoire de Chimie et Physique Quantique
UMR 5626 CNRS---Université Toulouse III-Paul Sabatier
118 Route de Narbonne, F-31062 Toulouse, France}

\author{Maen Salman} 
\affiliation{Laboratoire de Chimie et Physique Quantique
UMR 5626 CNRS---Université Toulouse III-Paul Sabatier
118 Route de Narbonne, F-31062 Toulouse, France}

\affiliation{Laboratoire Kastler Brossel, Sorbonne Universit\'e, CNRS, ENS-Universit\'e PSL,
Coll\'ege de France 4 place Jussieu, F-75005 Paris, France}

\author{Trond Saue} 
\affiliation{Laboratoire de Chimie et Physique Quantique
UMR 5626 CNRS---Université Toulouse III-Paul Sabatier
118 Route de Narbonne, F-31062 Toulouse, France}
\date{\today}

\begin{abstract}
\noindent
We report a method for the evaluation of the one-loop self-energy, to all orders in the external binding field, using a Gaussian basis set expansion. This choice of basis is motivated by its widespread use in molecular calculations. For a one-electron atom, our results show excellent agreement with those obtained using the exact Dirac--Coulomb wave functions. The developed method can be of interest for high-precision studies of heavy few-electron molecular systems, where the rigorous computation of QED corrections is currently a formidable task.

\end{abstract}
\maketitle
\section{Introduction}
\noindent
Gaussian-type basis functions were first introduced in molecular computations by Samuel Francis Boys in 1950 \cite{BoysI}. Although a single Gaussian function cannot accurately capture the wave function's asymptotic tail or its behavior near a point-like nucleus, the ease with which multicenter electron-electron repulsion integrals can be calculated greatly outweighs the need for a somewhat larger basis set expansion, as compared to, for instance, Slater-type functions. Consequently, Gaussian basis sets are almost universally used for the description of the electronic structure of polyatomic molecules \cite{HelgakerBook}. The rapid algorithmic development and the increase in computational resources have enabled the treatment of many-electron atoms and molecules with remarkable accuracy. To further reduce theoretical uncertainties, it is necessary to account for relativistic and quantum electrodynamics (QED) effects. Recently, precision measurements, carried out with atoms and molecules, combined with accurate theoretical predictions, have been proposed as tests of fundamental physics and means to search for potential new physics beyond the Standard Model \cite{Safronova18}. This line of research has raised an increasing interest in the rigorous computation of QED corrections to the energy levels and properties of molecular systems. Interestingly, George G. Hall noted about Boys `\emph{He [Boys] even tried to bring Gaussians into quantum field theory. ... Nothing came of it but he was not disheartened}' ~\cite{Hall96}. We have been unable to find further details about these ideas.

The leading-order QED corrections to the energy levels of a one-electron atom are vacuum polarization and self-energy. Recently the non-linear vacuum polarization density was considered using a Gaussian basis set by the two of us \cite{SaSa23}, and independently in Ref.~\cite{IvBaGlVo24}, where the shift of the energy levels was also evaluated. The treatment of the self-energy correction to all orders in the Coulomb field of the nucleus is essential to the accurate theoretical determination of energy levels and properties of heavy, few-electron atoms. The first calculation was carried out by Desiderio and Johnson \cite{DeJo71} using the method proposed by Brown, Langer and Schaefer \cite{BrLaSc59}. Later, methods for precise evaluation were pioneered by Mohr \cite{Mohr74,Mohr74-2} and further developed and extended to excited states, finite nuclei, and low nuclear charge \cite{Mohr75,Mohr82,MoKi92,Mohr92,InMo92,MoSo93,InMo98,JeMoSo99}. For a detailed review of this approach see, Ref.~\cite{MoPlSo98}. The high-precision evaluation usually requires the use of exact Dirac--Coulomb wave functions, summing over intermediate states with high angular momentum and principal quantum number, which restricts the applicability of this method to simple atomic systems. The first self-energy calculation, using a finite basis to describe the Dirac--Coulomb wave functions, was carried out by Blundell and Snyderman \cite{BlundellSnyderman91}, using $B$-spline basis functions and the many-potential expansion (\textit{vide infra}) proposed by Snyderman \cite{Snyderman91}. The use of the basis set expansion allowed for the generalization of their method to arbitrary local potentials, including both finite nuclear size and screening effects \cite{Blundell92}. The first application of the $B$-spline method for the evaluation of the self-energy of a one-electron diatomic quasimolecule was reported in Ref.~\cite{ArSu15}. As an alternative way to handle the divergences, the partial-wave renormalization was developed independently by Lindgren, Persson and Salomonson \cite{PeLiSa93,LiPeSaYn93} and by Quiney and Grant \cite{QuGr93,QuGr94}. This approach was recently proposed to be suitable for numerical computation of QED corrections in few-electron systems \cite{MaMa24}. For a recent review on computational methods of QED effects see, Ref.~\cite{YeMa20}.

In this Letter, we present a method for the evaluation of the one-loop self-energy correction to the energy levels of a one-electron atom using a Gaussian basis set expansion. This basis set allows for the efficient and high-precision description of molecular wave functions, enabling the generalization of the method to polyatomic systems. The developed approach is a significant advancement towards the rigorous evaluation of QED corrections in general molecules, aiding the high-precision tests of QED and the search for physics beyond the Standard Model. We use full SI units throughout this work.

\section{Dirac equation in a radial potential}
The Dirac equation in an external potential is
\begin{align}
    \left( \beta mc^2 -\iim \hbar c\balpha\cdot \grad + V(r)\right) \psi_{a\kappa}(\br) = E_{a\kappa}\psi_{a\kappa}(\br)  \, ,
    \label{eq:direq}
\end{align}
where $V(r)$ is the potential energy associated with a radial scalar potential. In the present work, it was chosen as the Coulomb potential of a point-like nucleus $V(r)=-Ze^2/(4\pi\varepsilon_0 r)$, though the method is readily extendable to finite nuclear models or screened potentials for the case of many-electron systems. The wave function can be written as a product of radial and angular solutions
\begin{align}
    \psi_{a\kappa m_{j}}(\br)=&\frac{1}{r}
    \begin{pmatrix}P_{a\kappa}(r)\chi_{\kappa m_{j}}(\hat{\boldsymbol{r}})\\
    \mathrm{i}Q_{a\kappa}(r)\chi_{-\kappa m_{j}}(\hat{\boldsymbol{r}})
    \end{pmatrix} \, .
    \label{eq:wf}
\end{align} 
Here, the $a$ index enumerates the radial solutions, $\kappa$ is the quantum number specifying both orbital- and total-angular momenta, and $m_j$ is the projection of the total angular momentum onto the $z$ axis. The spin-angular momentum eigenfunctions are given as
\begin{align}
    \chi_{\kappa m_{j}}(\hat{\boldsymbol{r}})=&\sum_{m_{\ell},m_{s}}\left\langle \ell,m_{\ell},1/2,m_{s}\left|j,m_{j}\right.\right\rangle Y_{\ell}^{m_{\ell}}(\hat{\boldsymbol{r}})\boldsymbol{u}_{m_{s}} \, ,
\end{align}
where $\left\langle \ell,m_{\ell},1/2,m_{s}\left|j,m_{j}\right.\right\rangle$ are Clebsch--Gordan coefficients, $Y_{\ell}^{m_{\ell}}(\hat{\boldsymbol{r}})$ are spherical harmonics and  $\boldsymbol{u}_{1/2}=(1,0)^T$, $\boldsymbol{u}_{-1/2}=(0,1)^T$. Inserting Eq.~\eqref{eq:wf} into the Dirac equation we eliminate the angular variables and obtain the radial Dirac equation
\begin{align}
    \begin{pmatrix}
        V(r)+mc^2 & 
        -c\hbar\left( \frac{\dd}{\dd r}-\frac{\kappa}{r}\right) \\
        c\hbar\left( \frac{\dd}{\dd r}
        +\frac{\kappa}{r}\right) & 
        V(r)-mc^2
    \end{pmatrix} 
    \begin{pmatrix} 
        P_{a\kappa}(r) \\
        Q_{a\kappa}(r) 
    \end{pmatrix}
    = E_{a\kappa} 
    \begin{pmatrix} 
        P_{a\kappa}(r) \\
        Q_{a\kappa}(r) 
    \end{pmatrix} \, .
\end{align}
The radial solutions are expressed as linear combinations of $\nb$ basis functions
\begin{align}
    P_{a\kappa}(r) = \sum_i^{\nb} c_{ia\kappa}^P g^P_{\kappa i}(r) 
    &&
    Q_{a\kappa}(r) = \sum_i^{\nb} c_{ia\kappa}^Q g^Q_{\kappa i}(r) \, .
\end{align}
The radial basis functions are chosen as normalized Gaussian functions, multiplied by an appropriate power of $r$ to capture the short-range behavior of the exact wave function with a finite nuclear charge distribution
\begin{align}
    g_{\kappa i}^{P}(r)&= \mathcal{N}_{\kappa}(\zeta_i)r^{\left|\kappa+\frac{1}{2}\right|+\frac{1}{2}}\mathrm{e}^{-\zeta_{i}r^{2}} \, ,
    \label{eq:Gaussian}
\end{align}
where $\mathcal{N}_\kappa (\zeta_i)$ is a normalization factor. The large- and small component basis sets are connected by the restricted kinetic balance condition \cite{StHa84}
\begin{align}
    g^Q_{\kappa i}(r) = \frac{\hbar}{2mc}
    \left( \frac{\dd}{\dd r} + \frac{\kappa}{r} \right) g^P_{\kappa i}(r) \, .
\end{align}
This construction is significantly simpler than the dual kinetic balance approach, which was first introduced for self-energy calculations using $B$-spline basis in Ref.~\cite{ShTuYePlSo04}. While both approaches are strictly valid only for non-singular potentials, in practice, Gaussian basis sets can usually be safely used with either restricted or dual kinetic balance, even for a point-like nucleus \cite{SaSa20,SaSa23}. The matrix representation of the Hamiltonian and overlap can be constructed analytically, and the eigenvalues and linear expansion coefficients are obtained numerically. The explicit form of the matrix elements is given in Ref.~\cite{SaSa20}.

\section{One-loop self-energy}

\begin{figure}
    \centering
    \includegraphics[width=\linewidth]{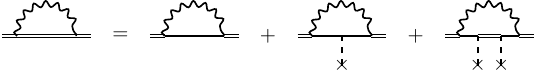}
    \caption{
    A diagrammatic representation of the many-potential expansion of the bound self-energy. Double internal lines represent the bound-electron propagator, single lines indicate the free-electron propagator, wavy lines correspond to virtual photons, and dashed lines terminated by a cross depict interactions with the Coulomb potential of the nucleus.
    }
    \label{fig:mpe}
\end{figure}
The shift in the energy levels of a one-electron atom, due to the one-loop self-energy correction is given by the real part of the expression \cite{BaBeFe53,schweber1961}
\begin{align}
    \Delta E_a^\text{SE}=&
    2\alpha\varepsilon_0 c^2 \int_{C_\text{F}}\text{d}z
    \int\text{d}^{3}\boldsymbol{x}_{1}
    \int\text{d}^{3}\boldsymbol{x}_{2} \,
    D(z,\boldsymbol{x}_{12})\psi_{a}^{\dagger}(\boldsymbol{x}_{1})\alpha_{\mu}G\left(E_{a}-z,\boldsymbol{x}_{1},\boldsymbol{x}_{2}\right)\alpha^{\mu}\psi_{a}(\boldsymbol{x}_{2}) \nonumber \\
    &-\delta m c^2 \int\text{d}^{3}\boldsymbol{x}\,\psi_{a}^{\dagger}(\boldsymbol{x})\beta\psi_{a}(\boldsymbol{x})
    \, ,
    \label{eq:boundse}
\end{align}
where $\alpha^\mu=(I_4, \boldsymbol{\alpha})$ are the Dirac matrices, the integral over $z$ is taken along the Feynman contour $C_\text{F}$, $\delta m$ is the mass-renormalization counterterm, $G(z,\bx_1,\bx_2)$ is the coordinate-representation of the bound-electron Green function $\mathcal{G}(z)=\left(z-\mathcal{H} \right)^{-1}$, and $D(z,\bx)$ is the photon propagator in energy-coordinate representation, which in Feynman gauge is
\begin{align}
    D(z,\boldsymbol{x}_{12})
    =
    - \frac{\iim  \hbar}{c\varepsilon_0}  \int\frac{\dd^{3}\bk}{(2\pi)^{3}}\,\frac{\eem^{\iim\bk\cdot(\bx_{1}-\bx_{2})}}{(z/\hbar c )^2-\boldsymbol{k}^{2}+\iim\epsilon } \, .
    \label{eq:photonpropagator}
\end{align}
The main difficulty in the numerical evaluation of Eq.~\eqref{eq:boundse} as it stands is that it is formulated as the difference between two divergent terms. To cancel the divergence between the unrenormalized bound-electron self energy and the counterterm, we first expand the bound-electron propagator in powers of the external potential \cite{Snyderman91,YerokhinShabaev99}. 
In operator form
\begin{align}
    \mathcal{G}(z) =
    \frac{1}{z-\mathcal{H}} =
    \frac{1}{z-\mathcal{H}_0} 
    +
    \frac{1}{z-\mathcal{H}_0} \mathcal{V} \frac{1}{z-\mathcal{H}_0}
    +
    \frac{1}{z-\mathcal{H}_0} \mathcal{V} \frac{1}{z-\mathcal{H}} \mathcal{V} \frac{1}{z-\mathcal{H}_0} \,
    ,
\end{align}
where $\mathcal{H}_0$ is the free-electron Hamiltonian, $\mathcal{V}$ is the potential-energy operator of the nuclear attraction, and  $\mathcal{H}=\mathcal{H}_0 + \mathcal{V} $. By inserting the many-potential expansion into Eq.~\eqref{eq:boundse}, as shown diagrammatically in Fig.~\ref{fig:mpe}, rewriting the first two terms in momentum space, and isolating the ultraviolet divergence, we observe the explicit cancellation of the infinite terms and the mass counterterm \cite{Snyderman91,YerokhinShabaev99}. Hence, we are left with
\begin{align}
    \Delta E_a^\text{SE}=&
    \int \text{d}^3\bp\,
    \bar{\psi}_a(\bp) 
    \Sigma_\text{R}(\text{p})
    \psi_a(\bp) \nonumber \\
    &+\int \text{d}^3\bp_1\int \text{d}^3\bp_2\,
    \bar{\psi}_a(\bp_1) \Gamma_\text{R}^0(\text{p}_1,\text{p}_2)
    V(\left|\bp_1-\bp_2\right| )
    \psi_a(\bp_2) \nonumber \\
    &+\Delta E^\text{mp}_a \nonumber \\
    =& \Delta E^{(0)}_{a;\text{R}} + \Delta E^{(1)}_{a;\text{R}}  + \Delta E^\text{mp}_a \, ,
\end{align}
where $\bar{\psi}=\psi^\dagger\beta$ is the Dirac adjoint. The three terms are referred to as the zero-, one-, and many-potential terms, respectively. $\Sigma_\text{R}(\text{p})$ and $\Gamma_\text{R}^\mu(\text{p}_1,\text{p}_2)$ are the renormalized self-energy and vertex operators, respectively, in terms of 4-momentum $\text{p}$, with their explicit forms given for instance in Ref.~\cite{YerokhinShabaev99}. $V(|\bp_1-\bp_2|)$ is the external field in momentum space, which, for the Coulomb potential of a point nucleus is 
\begin{align}
    V(|\bp|) = 
    \int \frac{\dd^3 \br}{(2\pi\hbar)^3} \eem^{-\frac{\iim}{\hbar} \bp \cdot \br }
    V(r)
    = -\frac{Z e^2}{(2\pi)^3\varepsilon_0\hbar |\bp|^2}  \, .
\end{align}
To evaluate the integrals in the zero-and one-potential terms, we introduce the momentum-space wave function
\begin{align}
    \psi_{a\kappa m_j}(\boldsymbol{p})=&\int
    \frac{\text{d}^{3}\boldsymbol{r}}{(2\pi\hbar)^{3/2}}
    \, \text{e}^{-\frac{\iim}{\hbar}\boldsymbol{p}\cdot\boldsymbol{r}}\psi_{a\kappa m_{j}}(\boldsymbol{r})=
    (-\iim)^{\ell}\frac{1}{p} 
    \begin{pmatrix}
      \tilde{P}_{a\kappa }(p) \chi_{\kappa m_{j}}(\hat{\boldsymbol{p}}) \\
      \tilde{Q}_{a\kappa}(p) \chi_{-\kappa m_{j}} (\hat{\boldsymbol{p}})
    \end{pmatrix} \, ,
\end{align}
with radial components 
\begin{align}
    \tilde{P}_{a\kappa}(p)&=\sqrt{\frac{2}{\pi\hbar^3}}p \int_{0}^{\infty}\text{d}r\,rj_{\ell}(pr/\hbar)P_{a\kappa}(r) \\
    \tilde{Q}_{a\kappa}(p)&=-\text{sgn}(\kappa)  \sqrt{\frac{2}{\pi\hbar^3}}p\int_{0}^{\infty}\text{d}r\,rj_{\bar{\ell}}(pr/\hbar)Q_{a\kappa}(r) \, ,
\end{align}
where $\ell=\left|\kappa +1/2 \right|-1/2$, $\bar{\ell}=\left|\kappa -1/2\right|-1/2$ and $j_\ell(x)$ are spherical Bessel functions. The integral over the Gaussian basis functions can be carried out analytically yielding the simple expression (see, 6.631.4 of Ref.~\cite{gradshteyn2007})
\begin{align}
    \tilde{g}^P(p) =\sqrt{\frac{2}{\pi\hbar^3}}p\int_{0}^{\infty}\dd r\,rj_{\ell}(pr)g^P(r) =
    \mathcal{N}_{\kappa}(\tilde{\zeta}_i) p^{\left|\kappa+\frac{1}{2} \right|+\frac{1}{2}}\text{e}^{-\tilde{\zeta}_{i}p^{2}}
\end{align}
with $\tilde{\zeta}_{i}=\frac{1}{4\zeta_{i}\hbar^2}$. The small component of the momentum-space wave function can be readily constructed by applying the kinetic balance condition in momentum space, and using the identity (see, 7.2.5.23 of Ref.~\cite{VaMoKh})
\begin{align}
    (\boldsymbol{\sigma}\cdot\boldsymbol{p})
    \chi_{\kappa m_{j}}(\hat{\boldsymbol{p}}) = -p \chi_{-\kappa m_{j}}(\hat{\boldsymbol{p}}) \, .
\end{align}
The straightforward form of the Fourier-transformed wave function in the Gaussian basis allows for efficient evaluation of the zero- and one-potential terms in momentum space. The remaining integrals over momenta can then be carried out numerically. Detailed description of the numerical integration scheme can be found for instance in Refs.~\cite{Blundell92,YerokhinShabaev99}. 

The many-potential term is free from divergences. For its evaluation, we turn to the spectral representation of the Green function. For each $\kappa$ we define the bound-electron Green function as
\begin{align}
    G_\kappa (z,\boldsymbol{x}_{1},\boldsymbol{x}_{2}) = \sum_{n}\frac{\psi_{n\kappa}(\boldsymbol{x}_{1})\psi_{n\kappa}^{\dagger}(\boldsymbol{x}_{2})}{z - E_{n\kappa}} \; ,
    \label{eq:boundG}
\end{align}
where the summation is understood to run over all positive- and negative-energy bound and continuum solutions of Eq.~\eqref{eq:direq} with the given $\kappa$ quantum number. The many-potential term is evaluated by explicitly subtracting the unrenormalized zero- and one-vertex contributions from the bound self-energy term in coordinate space, with each term evaluated in the same finite-basis for each set of intermediate states with $\kappa$ quantum number, that is 
\begin{align}
    \Delta E_\kappa^\text{mp} = \Delta E_\kappa^\text{bound}
    -\Delta E_\kappa^{(0)}
    -\Delta E_\kappa^{(1)} \, .
    \label{eq:Emp}
\end{align}
We define the bound-term, $\Delta E_\kappa^\text{bound}$, as the first term in Eq.~\eqref{eq:boundse}, using $G_\kappa$ in place of $G$. The zero- and one-vertex terms, $\Delta E_\kappa^{(0)}$ and $\Delta E_\kappa^{(1)}$, are given by the same expression, except that the bound-electron Green function is replaced by the zero- and one-potential Green functions, $G^{(0)}_\kappa$ and $G^{(1)}_\kappa$, respectively. In each case, the sum over intermediate states is restricted to the same, fixed $\kappa$ value. The zero-potential Green function is obtained by replacing the bound-electron wave functions and energies with free-particle ones in Eq.~\eqref{eq:boundG}. The one-potential Green function is given in coordinate-space as the integral
\begin{align}
    G^{(1)}_\kappa (z,\boldsymbol{x}_{1},\boldsymbol{x}_{2}) =
    \int \dd^3 \by \, 
    G^{(0)}_\kappa (z,\boldsymbol{x}_{1},\by) 
    V(\by)
    G^{(0)}_\kappa (z,\by,\boldsymbol{x}_{2})  \, .
\end{align}
The free-electron eigenstates and corresponding energies are obtained using the same basis set expansion as for the bound states, but the nuclear charge is set to zero in the computation. Finally, the $\Delta E_\kappa^\text{mp}$ contributions for each $\kappa$ value are summed to yield the complete many-potential term $\Delta E^\text{mp}$. In practice, the evaluation is carried out up to some maximal $|\kappa|$ value, depending on the nuclear charge and the reference state, and then the remaining contribution to the sum is obtained by fitting a polynomial in $|\kappa|^{-1}$. 

The integrals over the angular degrees of freedom and sums over Clebsch--Gordan coefficients can be calculated using angular momentum algebra. The explicit expressions can be found for instance in Ref.~\cite{YeMa20}. All radial integrals over coordinates involving Gaussian basis functions were calculated analytically using 6.631.1 of Ref.~\cite{gradshteyn2007}. For each term, the integral over the Feynman contour and the angular integrals in Eq.~\eqref{eq:photonpropagator} are evaluated analytically first, whereas the subsequent integration over the length of the photon-momentum vector is carried out by quadrature \cite{PeLiSa93,LiPeSaYn93,QuGr93}. For excited states, the integral can be taken in the principal-value sense, when integrating over the poles that appear at bound-state energies, yet this case is not considered in the present work.

\section{Results and discussion}
The self-energy correction is often given in terms of the slowly varying and dimensionless function $F(Z\alpha)$ defined as \cite{Mohr74,Mohr74-2}
\begin{align}
    \Delta E^\text{SE} = \frac{\alpha}{\pi} \frac{(Z\alpha)^4}{n^3}F(Z\alpha) mc^2  \, ,
    \label{eq:FZa}
\end{align}
where $n$ is the principal quantum number of the reference state. We carried out the computation for the ground state ($1s_{1/2}$) of hydrogen-like uranium ($Z=92$). We used $\alpha^{-1} = 137.035\, 989\, 5$ to match the calculations of Ref.~\cite{YerokhinShabaev99}. We assumed a point-like nucleus throughout. The basis set exponents were chosen according to the even-tempered scheme, such that $\zeta_i = \zeta_1 \beta^{i-1} $ where $i=1,\ldots,\nb$. This construction is known to become complete in the limit $\zeta_1 \rightarrow 0$, $\beta \rightarrow 1^+ $, and $\nb \rightarrow \infty$ \cite{Klahn85}. In practical calculations however, the $\beta$ value can not be set too close to unity, as it leads to linear dependencies in the basis set. Nevertheless, the above limits can be used to systematically improve the basis set. In the present work the parameters were chosen as $\zeta_1=0.01~a_0^{-2}$, $\beta = 1.5$, and the number of basis functions was chosen as $\nb = 100$, which yielded a ground-state energy with relative precision better than $10^{-9}$. The method was implemented in \textsc{fortran90}, using standard double-precision arithmetic. For the diagonalization of the Hamiltonian the \textsc{dsygv} subroutine of the \textsc{lapack} library was used \cite{lapack}. No linear-dependence issues were observed with the above parametrization and algorithm, even when using exceedingly large basis sets. With careful implementation of the matrix elements, the finite-basis solution of the radial Dirac equation is possible even for states with $|\kappa | \sim 100 $ without need for increased numerical precision. Numerical integration was performed with the \textsc{quadpack} library subroutines \cite{quadpack83}. The many-potential term was evaluated up to $|\kappa|=15$ and the sum over $\kappa$ was extrapolated as described above. The computation of the bound and zero-potential terms for a single $\kappa$ value took about 1 hour, whereas the computation of the one-potential term is more involved and can take up to 1.5 days on an AMD 2.8~GHz processor. However, the one-potential Green function can be more efficiently computed by numerically evaluating the derivative of the bound-electron Green function $G^{(1)}_\kappa =Z\partial_Z G_\kappa\eval_{Z=0}$. With that, the computation of the one-vertex term has approximately the same cost as the bound contribution. Nevertheless, compared to the exact Dirac--Coulomb Green function method reported in Ref.~\cite{YerokhinShabaev99}, the basis set method remains computationally more intensive. In Table~\ref{tab:manypot}, the contributions to the many-potential term are shown for $\kappa=\pm 1$. None of the individual terms agree to better than three significant digits with the results of Ref.~\cite{YePriv24}, which were obtained using the exact Dirac--Coulomb Green function and the method described in Ref.~\cite{YerokhinShabaev99}. However, the many-potential term, computed as the difference defined in Eq.~\eqref{eq:Emp} is in perfect agreement. This remarkable cancellation of error allows for the accurate computation of the many-potential term in a Gaussian basis, even though this basis set does not seem to be suitable for the precise evaluation of either the zero-, one-vertex or bound terms. A detailed analysis on the origins of this cancellation will be given in a subsequent paper. The largest sources of uncertainty of the final result are the finite-basis error in the individual many-potential contributions and the uncertainty of the extrapolation in $|\kappa|$. The basis set error was estimated by repeating the computation for selected $\kappa$ values with different basis set parameters and assuming uniform error for each term. The convergence in the basis set is illustrated in Table~\ref{tab:convergence}, where the contribution from the $|\kappa|= 5$ intermediate states is shown with different basis set sizes and $\beta$ values. The $\zeta_1$ exponent determines the largest distance that can be represented within the basis set. Provided that it is sufficiently small to describe the long-range behavior of the wave functions, its variation has little influence on the final result. The obtained values converge monotonically to the reference value, yet the convergence rate with the number of basis functions soon becomes slow. The error of the extrapolation of the partial wave expansion was estimated by comparing the results obtained from different extrapolation schemes. The obtained final results for the zero- and one-potential terms in momentum space, and the complete many-potential term are presented in Table~\ref{tab:results}. The values are in perfect agreement with the exact Dirac--Coulomb Green function approach of Ref.~\cite{YerokhinShabaev99}. The many-potential contribution for each $\kappa$ value is given in Table~\ref{tab:fullresults}. 

In summary, we have developed a method for the calculation of the one-loop self-energy in a finite Gaussian basis set. To demonstrate its efficiency, we evaluated the self-energy correction for the ground state of hydrogen-like uranium, achieving perfect agreement with the results of the exact Dirac--Coulomb Green function method. The present approach allows for the use of arbitrary nuclear charge distributions and we expect that it will be also applicable for the evaluation of higher-order self-energy corrections. The use of Gaussian-type basis functions opens the way for the computation of rigorous QED corrections in molecular systems. 

\begin{table}
        \caption{Contributions to the many-potential term for the ground state of hydrogen-like uranium, given in terms of the function $F(Z\alpha)$ defined in Eq.~\eqref{eq:FZa}. The many-potential term is computed as the difference given in Eq.~\eqref{eq:Emp}. The last column shows the difference from the values obtained with the exact Dirac--Coulomb Green function method \cite{YePriv24}.}
        \label{tab:manypot}
        \begin{tabular}{c l r r r}
        \hline \hline
        \multicolumn{1}{c}{$\kappa$} & \multicolumn{1}{c}{Term}& \multicolumn{1}{c}{Present work} & \multicolumn{1}{c}{Ref.~\cite{YePriv24}} & \multicolumn{1}{c}{\ \ Diff.} \\
        \hline
        \multirow{4}{*}{$-1$}
        &$\Delta E_\kappa^{(0)}$	& $2.421\, 600\, 5$  & $2.426\, 503\, 43$ &	 $-0.004\, 902\, 9$\\
        &$\Delta E_\kappa^{(1)}$	& $0.338\, 788\, 8$	 & $0.336\, 957\, 11$ &	 $0.001\, 831\, 7$\\
        &$\Delta E_\kappa^\text{bound}$	& $4.271\, 705\, 9$	 & $4.274\, 776\, 90$ &	 $-0.003\, 071\, 0$\\
        \cline{2-5}
        &$\Delta E_\kappa^\text{mp}$	& $1.511\, 316\, 6$	 & $1.511\, 316\, 36$ &	 $0.000\, 000\, 2$\\
        \hline
        \multirow{4}{*}{$1$}
        &$\Delta E_\kappa^{(0)}$	&  $ 2.579\, 083\, 4$  & $ 2.584\, 060\, 05$ &  $-0.004\, 976\, 7$ \\
        &$\Delta E_\kappa^{(1)}$	&  $-0.502\, 067\, 9$  & $-0.503\, 886\, 64$ &  $0.001\, 818\, 8$\\
        &$\Delta E_\kappa^\text{bound}$ & $ 2.197\, 906\, 1$  &	$ 2.201\, 063\, 89$ &  $-0.003\, 157\, 7$\\
        \cline{2-5}
        & $\Delta E_\kappa^\text{mp}$	&  $ 0.120\, 890\, 7$  & $ 0.120\, 890\, 48$ &  $ 0.000\, 000\, 2$\\
        \hline \hline
        \end{tabular}
    \end{table}    

\begin{table}
    \centering
    \caption{Convergence of the $\Delta E_{|5|}^\text{mp} = \Delta E_{-5}^\text{mp}+\Delta E_{5}^\text{mp}$ partial-wave contribution to the many-potential term with respect to the basis set parameters $\nb$ and $\beta$ for the ground state of hydrogen-like uranium. All values are given in terms of $F(Z\alpha)$, defined in Eq.~\eqref{eq:FZa}. The smallest exponent was fixed to $\zeta_1=0.01~a_0^{-2}$. The difference is given with respect to the reference value $\Delta E_{|5|}^\text{mp,Ref}=0.001\ 988\ 36$, obtained with the exact Dirac--Coulomb Green function method in Ref.~\cite{YePriv24} (see also Ref.~\cite{Yerokhin05}).}
    
    \label{tab:convergence}
    \begin{tabular}{@{} c@{\ } d{2.10}@{\ \ } d{2.10}@{\ \ } d{2.10}@{\ \ } d{2.10}@{\ \ } d{2.10}@{\ \ } d{2.10} @{}}
        \hline \hline
        \multicolumn{1}{c}{}& 
        \multicolumn{2}{c}{$\beta=1.50$} & 
        \multicolumn{2}{c}{$\beta=1.45$} &
        \multicolumn{2}{c}{$\beta=1.40$}  \\
        \cline{2-7}
        \multicolumn{1}{c}{$\nb $}& 
        \multicolumn{1}{c}{$\Delta E_{|5|}^\text{mp}$} & 
        \multicolumn{1}{c}{Diff.} &
        \multicolumn{1}{c}{$\Delta E_{|5|}^\text{mp}$} & 
        \multicolumn{1}{c}{Diff.} &
        \multicolumn{1}{c}{$\Delta E_{|5|}^\text{mp}$} & 
        \multicolumn{1}{c}{Diff.} \\
        \hline
        40	&	0.001\ 700\ 10	&	-0.000\ 288\ 25	&	0.001\ 366\ 56	&	-0.000\ 621\ 79	&	0.000\ 824\ 58	&	-0.001\ 163\ 77	\\
        50	&	0.001\ 972\ 56	&	-0.000\ 015\ 79	&	0.001\ 933\ 82	&	-0.000\ 054\ 53	&	0.001\ 813\ 15	&	-0.000\ 175\ 20	\\
        60	&	0.001\ 989\ 22	&	 0.000\ 000\ 87	&	0.001\ 985\ 66	&	-0.000\ 002\ 69	&	0.001\ 972\ 48	&	-0.000\ 015\ 87	\\
        70	&	0.001\ 990\ 08	&	 0.000\ 001\ 73	&	0.001\ 989\ 26	&	 0.000\ 000\ 91	&	0.001\ 987\ 60	&	-0.000\ 000\ 75	\\
        80	&	0.001\ 990\ 13	&	 0.000\ 001\ 78	&	0.001\ 989\ 50	&	 0.000\ 001\ 15	&	0.001\ 988\ 89	&	 0.000\ 000\ 54	\\
        90	&	0.001\ 990\ 12	&	 0.000\ 001\ 77	&	0.001\ 989\ 51	&	 0.000\ 001\ 16	&	0.001\ 989\ 08	&	 0.000\ 000\ 73	\\
        \hline \hline
    \end{tabular}

\end{table}

\begin{table}
    \centering
    \caption{Contributions to the total self-energy correction for the ground state of hydrogen-like uranium in terms of $F(Z\alpha)$ defined in Eq.~\eqref{eq:FZa}. }
    \label{tab:results}
        \begin{tabular}{@{} c@{\ \ \ } d{2.8}@{\ \ \ } d{2.8}@{\ }  @{}}
        \hline \hline
         \multicolumn{1}{c}{Term}& \multicolumn{1}{c}{Present work} & \multicolumn{1}{c}{Ref.~\cite{YerokhinShabaev99}}  \\
        \hline
        $\Delta E^{(0)}_\text{R}$ & -2.155\ 55 &  -2.155\ 55  \\
        $\Delta E^{(1)}_\text{R}$ &  1.984\ 01 &   1.984\ 01  \\
        $ \Delta E^\text{mp} $
        & 1.662\ 44(5)  & 1.662\ 45(2)  \\
        $ \Delta E^\text{SE} $
        & 1.490\ 89(5) & 1.490\ 91(2)  \\
        \hline \hline
        \end{tabular}
\end{table}

\section{Acknowledgement}
DF and TS thank Vladimir Yerokhin for the many helpful discussions and for sharing his detailed results with us. This project was funded by the European Research Council (ERC) under the European Union’s Horizon 2020 research and innovation programme (Grant Agreement No. ID:101019907).

\begin{table}[H]
    \centering
    \caption{Contributions to the total self-energy correction for the ground state of hydrogen-like uranium, given in terms of the function $F(Z\alpha)$ defined in Eq.~\eqref{eq:FZa}. The many-potential term is computed as the difference given in Eq.~\eqref{eq:Emp}. The estimated error of the many-potential term is in the last two digits.}
    \label{tab:fullresults}
        \begin{tabular}{@{} r @{\ \ } d{2.10}@{\ \ } d{2.10}@{\ \ } d{2.10}@{\ \ } d{2.10}  @{}}
        \hline \hline
         \multicolumn{1}{c}{$\kappa$}& 
         \multicolumn{1}{c}{$\Delta E_\kappa^{(0)}$} & 
         \multicolumn{1}{c}{$\Delta E_\kappa^{(1)}$} &
         \multicolumn{1}{c}{$\Delta E_\kappa^\text{bound}$} &
         \multicolumn{1}{c}{$\Delta E_\kappa^\text{mp}$}  \\
        \hline
$-1	$ &	2.421\ 600\ 51	&	 0.338\ 788\ 80	&	4.271\ 705\ 88	&	1.511\ 316\ 58	\\
$1	$ &	2.579\ 083\ 35	&	-0.502\ 067\ 88	&	2.197\ 906\ 14	&	0.120\ 890\ 67	\\
$-2	$ &	1.125\ 020\ 84	&	-0.271\ 112\ 05	&	0.846\ 791\ 21	&  -0.007\ 117\ 58	\\
$2	$ &	1.587\ 874\ 79	&	-0.293\ 652\ 31	&	1.313\ 383\ 12	&	0.019\ 160\ 63	\\
$-3	$ &	0.882\ 387\ 35	&	-0.201\ 185\ 00	&	0.683\ 565\ 45	&	0.002\ 363\ 10	\\
$3	$ &	1.122\ 218\ 52	&	-0.211\ 584\ 93	&	0.916\ 585\ 15	&	0.005\ 951\ 55	\\
$-4	$ &	0.707\ 952\ 12	&	-0.159\ 434\ 80	&	0.549\ 822\ 64	&	0.001\ 305\ 32	\\
$4	$ &	0.855\ 167\ 57	&	-0.163\ 810\ 81	&	0.693\ 858\ 86	&	0.002\ 502\ 09	\\
$-5	$ &	0.584\ 171\ 13	&	-0.130\ 043\ 78	&	0.454\ 857\ 84	&	0.000\ 730\ 48	\\
$5	$ &	0.683\ 200\ 93	&	-0.132\ 112\ 18	&	0.552\ 348\ 39	&	0.001\ 259\ 64	\\
$-6	$ &	0.492\ 812\ 60	&	-0.108\ 419\ 15	&	0.384\ 838\ 04	&	0.000\ 444\ 59	\\
$6	$ &	0.563\ 537\ 92	&	-0.109\ 499\ 86	&	0.454\ 753\ 08	&	0.000\ 715\ 03	\\
$-7	$ &	0.422\ 836\ 59	&	-0.091\ 959\ 43	&	0.331\ 167\ 21	&	0.000\ 290\ 05	\\
$7	$ &	0.475\ 536\ 59	&	-0.092\ 567\ 58	&	0.383\ 411\ 04	&	0.000\ 442\ 04	\\
$-8	$ &	0.367\ 588\ 90	&	-0.079\ 075\ 33	&	0.288\ 713\ 21	&	0.000\ 199\ 63	\\
$8	$ &	0.408\ 114\ 62	&	-0.079\ 435\ 71	&	0.328\ 970\ 07	&	0.000\ 291\ 16	\\
$-9	$ &	0.322\ 901\ 21	&	-0.068\ 757\ 47	&	0.254\ 286\ 96	&	0.000\ 143\ 22	\\
$9	$ &	0.354\ 828\ 27	&	-0.068\ 978\ 18	&	0.286\ 051\ 42	&	0.000\ 201\ 32	\\
$-10$ &	0.286\ 050\ 56	&	-0.060\ 340\ 14	&	0.225\ 816\ 58	&	0.000\ 106\ 16	\\
$10	$ &	0.311\ 688\ 56	&	-0.060\ 477\ 57	&	0.251\ 355\ 60	&	0.000\ 144\ 61	\\
$-11$ &	0.255\ 187\ 86	&	-0.053\ 368\ 28	&	0.201\ 900\ 35	&	0.000\ 080\ 77	\\
$11	$ &	0.276\ 096\ 48	&	-0.053\ 453\ 93	&	0.222\ 749\ 66	&	0.000\ 107\ 11	\\
$-12$ &	0.229\ 013\ 08	&	-0.047\ 520\ 61	&	0.181\ 555\ 25	&	0.000\ 062\ 79	\\
$12	$ &	0.246\ 284\ 64	&	-0.047\ 573\ 06	&	0.198\ 792\ 90	&	0.000\ 081\ 32	\\
$-13$ &	0.206\ 583\ 71	&	-0.042\ 563\ 74	&	0.164\ 069\ 63	&	0.000\ 049\ 67	\\
$13	$ &	0.221\ 006\ 48	&	-0.042\ 594\ 48	&	0.178\ 475\ 03	&	0.000\ 063\ 03	\\
$-14$ &	0.187\ 197\ 74	&	-0.038\ 323\ 91	&	0.148\ 913\ 71	&	0.000\ 039\ 88	\\
$14	$ &	0.199\ 354\ 91	&	-0.038\ 340\ 28	&	0.161\ 064\ 33	&	0.000\ 049\ 71	\\
$-15$ &	0.170\ 319\ 82	&	-0.034\ 668\ 96	&	0.135\ 683\ 29	&	0.000\ 032\ 42	\\
$15	$ &	0.180\ 651\ 88	&	-0.034\ 675\ 78	&	0.146\ 015\ 88	&	0.000\ 039\ 77	\\
        \hline \hline
        \end{tabular}
        \label{tab:allk}
\end{table}

\newpage
\bibliography{references}
\end{document}